\documentclass[a4paper,fleqn,usenatbib]{mnras}

\usepackage[T1]{fontenc}
\usepackage{ae,aecompl}

\usepackage{graphicx}	
\usepackage{amsmath}	
\usepackage{amssymb}	
\usepackage{subcaption}
\title[Accreted Star Cluster Dynamics]{The Dynamical Evolution of Accreted Star Clusters in the Milky Way}

\author[Miholics, Webb \& Sills]{
Meghan Miholics,\thanks{E-mail: miholim@mcmaster.ca (MM)}
Jeremy J. Webb,
and Alison Sills
\\
Department of Physics \& Astronomy, McMaster University, 1280 Main St. W., Hamilton, Ontario, L8S 4L8, Canada
}

\date{Accepted XXX. Received YYY; in original form ZZZ}

\pubyear{2015}

\begin{document}
\label{firstpage}
\pagerange{\pageref{firstpage}--\pageref{lastpage}}
\maketitle

\begin{abstract}
We perform $N$-body simulations of star clusters in time-dependant galactic potentials. Since the Milky Way was built-up through mergers with dwarf galaxies, its globular cluster population is made up of clusters formed both during the initial collapse of the Galaxy and in dwarf galaxies that were later accreted. Throughout a dwarf-Milky Way merger, dwarf galaxy clusters are subject to a changing galactic potential. Building on our previous work, we investigate how this changing galactic potential affects the evolution of a cluster's half mass radius. In particular, we simulate clusters on circular orbits around a dwarf galaxy that either falls into the Milky Way or evaporates as it orbits the Milky Way. We find that the dynamical evolution of a star cluster is determined by whichever galaxy has the strongest tidal field at the position of the cluster. Thus, clusters entering the Milky Way undergo changes in size as the Milky Way tidal field becomes stronger and that of the dwarf diminishes. We find that ultimately accreted clusters quickly become the same size as a cluster born in the Milky Way on the same orbit. Assuming their initial sizes are similar, clusters born in the Galaxy and those that are accreted cannot be separated based on their current size alone. 
\end{abstract}



\section{Introduction}

Globular clusters are giant groups of stars in bound spherical configurations, that reside in all types of galaxies. Most globular clusters formed in the early Universe, coeval with their parent galaxies \citep{Krauss03}. Hence, globular cluster populations trace the hierarchal formation of structure in the Universe, in which large galaxies are formed through the merging of smaller galaxy building blocks \citep{White78}. Each galaxy's globular cluster population will be made up of both clusters that were formed in the initial collapse of the galaxy as well as clusters that were formed elsewhere and were later accreted onto the galaxy through galaxy mergers \citep{Searle78}. In particular, the globular cluster population in the Milky Way is composed of clusters formed in the Milky Way and clusters that were formed in dwarf galaxies that have merged with the Milky Way in the past. \\

Observational evidence for two distinct populations of globular clusters in the Milky Way is abundant. Perhaps the most compelling evidence for this scenario comes from the dwarf galaxy Sagittarius which is merging with the Milky Way at present day \citep{Ibata94}. Several of the Galaxy's globular clusters (anywhere from 5-9) have been shown to be coherent with the Sagittarius stream in position and velocity space and thus can be associated with the original dwarf galaxy \citep[e.g.][]{DaCosta95,Palma02,Bellazzini03,Law10}. Additional evidence comes from examining the present day properties of Milky Way globular clusters. \citet{Zinn93} was the first to note that the Galaxy's clusters are divided into two groups that have distinct horizontal branch morphologies, kinematics and spatial distributions from one another. More recent work has shown that certain subsets of Milky Way clusters have similar horizontal branch morphologies, core radii and relationships between age and metallicity as clusters found in Milky Way satellites such as  the LMC, SMC, Fornax and Sagittarius \citep[e.g][]{Mackey04,Marin09,Forbes10,Leaman13}. These studies typically find about 25-35 per cent of clusters in the Milky Way have properties similar to clusters observed in dwarf galaxies, indicating that as many as 40-50 of the Galaxy's clusters were accreted from dwarf galaxies. These estimates suggest that the Milky Way has undergone $\approx$ 6-8 mergers with galaxies like the Sagittarius dwarf over its history. Evidence for accretion of clusters onto one galaxy from another exists in many other systems such as Andromeda and in larger systems such as galaxy clusters \citep{Collins09,Mackey14,DAbrusco15}.\\

Although it is well established that the Milky Way is host to a population of accreted clusters, little work has been done so far to determine how the changing Galactic environment affects these clusters. The dynamical evolution of globular clusters in static potentials has been studied extensively and is significantly affected by the galaxy in which the cluster lives \citep[e.g.][]{Baumgardt03,Webb14a,Webb14b}. Clusters are subject to two body relaxation as well as tidal forces from the host galaxy. Two body relaxation alters the orbits of stars through their dynamical interactions with one another and ultimately causes the outer layers of the cluster to expand. The difference between the force from the galaxy at the cluster's centre and the force from the galaxy acting at some other position in the cluster creates tidal forces on the stars in the cluster. The points where the tidal forces from the galaxy exactly balance the force from the cluster are known as the Lagrange points. The first and second Lagrange points lie along the axis through the galaxy centre and the cluster. Stars in general escape the cluster through one of these two points. Hence, the tidal forces set a rough maximum size for the cluster as it expands (known as the Jacobi or tidal radius, $r_j$) which is the distance between the cluster centre and the first Lagrange point. Outside of this radius stars feel a stronger gravitational force from the galaxy than the cluster and are thus stripped away from the cluster. For clusters in spherically symmetric potentials (or in axisymmetric potentials on circular orbits in the plane of symmetry), the Jacobi radius is given by:
\begin{equation}
\label{eq:Jacobi}
r_j = R_G\bigg( \frac{M_c}{3M_G}\bigg)^{1/3}
\end{equation}

\noindent where $R_G$ is the radius of the circular orbit, $M_c$ is the cluster's mass and $M_G$ is the mass of the galaxy enclosed by the cluster's orbit. For a more detailed discussion of the above ideas see \citet{Renaud11}.

 The actual physical size of a cluster can be smaller or approximately equal to this radius. If a cluster occupies a large majority of the volume set by the tidal radius it is said to be tidally filling, if not it is underfilling. The ratio of the cluster's half light radius (radius which contains half the light) to the Jacobi radius, $r_h/r_j$, is used to measure the degree to which a cluster fills its tidal radius and is referred to as the tidal filling factor. It is also useful to compare $r_j$ to the 95 or 99 percent Lagrange radii (radius that contains 95 or 99 per cent of the mass) to obtain an idea of how close the outer region of the cluster is to $r_j$. A cluster's tidal filling factor can have a great influence on its evolution since clusters that are more tidally filling are more susceptible to the tidal forces of the galaxy and will experience higher mass loss rates than clusters that are underfilling.

The evolution of globular clusters has been studied in a variety of static galactic potentials \citep[e.g.][]{Baumgardt03,Webb14a,Webb14b} as well as some semi-static potentials where changes in the potential are made instantaneously \citep{Madrid14,Miholics14}. However, given that star clusters are embedded in ever evolving environments, it is important to be able to study clusters in time dependant potentials. Such a task has been made possible with the recent inception of  \textsc{Nbody6tt}, an extension of the $N$-body code \textsc{Nbody6}  \citep{Renaud11,Renaud15a}. \textsc{Nbody6tt} allows the user to evolve stellar clusters in arbitrary time dependant galactic potentials and has been used to study the evolution of clusters in major galaxy mergers \citep{Renaud13} as well as clusters embedded in slowly growing dark matter haloes \citep{Renaud15b}. Similar techniques have also been implemented by \citet{Rieder13} who studied clusters in the tidal field extracted from a cosmological dark matter simulation. 

Recently, steps have been taken towards understanding the evolution of dwarf galaxy globular clusters that undergo a change in galactic potential. \citet{Bianchini15} simulated clusters in the centre of a dwarf galaxy potential that is instantaneously or slowly removed. They found that clusters expand in response to the changing potential but never become as extended as they would if they evolved solely in isolation. In our previous work, \citet{Miholics14}, we studied dwarf galaxy clusters in a similar context. We simulated clusters undergoing an instantaneous change in galactic potential, from that of a dwarf galaxy to the Milky Way, to understand the ultimate evolution of a cluster that has been brought into the Milky Way by a merger with a dwarf. We found that a cluster's size will adjust rapidly in response to the new galactic potential until it is the same size as a Milky Way cluster on that orbit. In our current work, we simulate clusters in idealized time dependant potentials representative of a dwarf-Milky Way merger. This method allows us to study the evolution of a cluster throughout the whole merger process rather than just before and after. We start all of our simulations by placing the cluster on a circular orbit around a dwarf galaxy that does one of two things: falls into the Milky Way or evaporates as it orbits around the Milky Way. We always keep the Milky Way potential fixed but explore the effects of varying the cluster's size as well as the mass of the dwarf galaxy.

\section{Methods}

\subsection{The Combined Galactic Potential}

We simulate clusters with the $N$-body code \textsc{Nbody6} \citep{Aarseth01,Aarseth03,Nitadori12}. Traditionally, \textsc{Nbody6} is able to evolve clusters under the gravitational influence of a single galaxy. However, to study the evolution of star clusters in a dwarf-Milky Way merger, we need to simulate them under the combined influence of both the dwarf galaxy and Milky Way. To accomplish this goal, we utilize the extension \textsc{Nbody6tt} \citep{Renaud11,Renaud15a} which allows for integration of star clusters in arbitrary galactic potentials. \textsc{Nbody6tt} offers two possible methods for implementing the galactic potential. The option which we utilize here uses an expansion of the galactic potential which yields the force on a cluster star (as a function of position with respect to the cluster centre):
\begin{equation}
\nabla\phi_G(\mathbf{r})  = \nabla \phi_G (\mathbf{0}) - \mathbf{T_t}(\mathbf{r})\cdot\mathbf{r} + O(\mathbf{r}^2)
\end{equation}

\noindent where $\mathbf{r}$ is the position of the star in the cluster, $\phi_G$ is the galactic potential and $\mathbf{T_t}$ is a tensor (referred to as the tidal tensor) given by the following: 
\begin{equation}
\mathbf{T_t}^{ij} (\mathbf{r}) = \bigg( - \frac{\partial^2 \phi_G}{\partial x^i \partial x^j}\bigg)_\mathbf{r}
\end{equation}

\noindent (for more details on this expansion see \citep{Renaud11}). To use this method, a series of tidal tensors, evaluated at the cluster's position within the potential, must be supplied to the code at discrete timesteps. The position of the cluster within the potential as a function of time, i.e. the cluster's orbit, is completely specified by the user. This method allows for the simple calculation of the tidal tensor from two galaxies since it is simply the summation of the tidal tensors for each individual galaxy (evaluated at the cluster's position with respect to that galaxy). Note, however, that this summation is a vector summation. The effects of the two tidal field strengths are, in practice, not additive since the points where the first galaxy's tidal field is the strongest (i.e. the Lagrange points, see Section 1) are not located at the same position around the cluster as the points where the second galaxy's tidal field is the strongest. 

To emulate the potential a cluster would feel during a dwarf-Milky Way merger, we simulate clusters in two main scenarios. In the first scenario, the cluster begins its evolution on a circular orbit around a point mass dwarf galaxy which falls into the Milky Way. The initial separation between the galaxies' centres is always set to 50 kpc. Starting at 3 Gyr, the separation between the two galaxies is decreased at a constant rate (10 kpc/750 Myr) until the dwarf galaxy reaches a certain distance from the Milky Way centre and stops. We chose to decrease the separation at a constant rate in order to keep the comparison of the tidal forces from each galaxy simple. The rate was chosen such that the cluster falls into the Milky Way relatively slowly compared to its orbit around the dwarf. We keep the cluster on a circular orbit around the dwarf galaxy as the dwarf falls in. This method allows us to directly study how the varying tidal field strength of the Milky Way affects the cluster's evolution while maintaining a constant tidal field strength for the dwarf.

In the second scenario, the cluster is simulated on a circular orbit around a point mass dwarf galaxy that evaporates over time. Initially the cluster orbits around the dwarf galaxy which in turn executes a circular orbit around the Milky Way. After 3 Gyr of evolution in this combined system, we decrease the mass of the dwarf according to the following equation:
\begin{equation}
M_D(t) = M_oe^{-6(t-3.0)}
\end{equation}

\noindent where t is the time in Gyr and $M_o$ is the original mass. As the mass decreases, we also decrease the radius of the cluster's orbit around the dwarf according to the following:
\begin{equation}
R (t) = R_oe^{-(t-3.0)}
\end{equation}

\noindent where t is again the time in Gyr and $R_o$ is the original radius. The functional form for $M_D (t)$ was chosen such that the mass of the dwarf is effectively zero with respect to the cluster's mass after 3.0 Gyr. $R(t)$ was then chosen such that the $r_j$ in the dwarf potential only increases as a function of time. Ultimately, the cluster is left orbiting around the Milky Way on a circular orbit at the same distance that the dwarf galaxy was from the centre. Although we choose the orbit of the cluster, this method allows us to study how the diminishing tidal field strength of the dwarf affects the evolution of the cluster while keeping the average tidal field strength of the Milky Way constant. 

The scenarios described above probe the two key processes in a dwarf-Milky Way merger that will affect the cluster; the increase in tidal field strength of the Milky Way as the dwarf falls into the galaxy and the decrease in the tidal field strength of the dwarf as it is stripped by the Milky Way. The relative contributions of these two processes will be determined by the realistic orbit of the cluster in the combined potential. However, combining the results from these two scenarios will allow us to obtain a full picture of a star cluster's evolution in a dwarf-Milky Way galaxy merger. 

\subsection{Simulations and Parameters}

All clusters are simulated with $N = 50,000$ stars distributed according to a Plummer density profile \citep{Plummer11} with no primordial binaries. Velocities are assigned such that the cluster is initially in virial equilibrium. Initial masses are assigned using a \citet{Kroupa01} initial mass function with masses from $0.1$ to $50 M_{\odot}$ and an average mass of $0.6 M_{\odot}$. The effects of stellar and binary evolution are implemented throughout the simulation as per the prescriptions in \citet{Tout97,Hurley00,Hurley08}. To characterize the actual size of a cluster, we use the half mass radius (radius that contains half the mass), $r_m$, since it is a proxy for the half light radius measured by observers. For most of our simulations, we set the initial $r_m$ equal to 3.2 pc, unless otherwise stated.

The clusters are simulated in the two scenarios described above which we will call for simplicity ``dwarf falls" and ``dwarf evaporates" for the dwarf falling into the Milky Way and the dwarf losing mass, respectively. In all simulations, the Milky Way is modelled as a point mass bulge, \citet{Miyamoto75} disc and logarithmic halo, details of which can be found in \citet{Miholics14}. For the base case in the dwarf falls scenario, we simulate a cluster on a $R = 4.0$ kpc circular orbit around a $10^9 M_{\odot}$ dwarf. The initial separation between the two galaxies is set to 50 kpc. We allow the dwarf to fall into the Milky Way until it reaches a distance of 15 kpc from the Milky Way's centre. We also examine the effect of changing the initial half mass radius by simulating a cluster in the same potential but with an initial $r_m = 4.0 pc$. Additionally, we investigate the effect of varying the dwarf's tidal field strength by performing a simulation with the mass of the dwarf, $M_D = 10^{10} M_{\odot}$. In the dwarf evaporates scenario, we simulate the cluster around a $10^10 M_{\odot}$ dwarf with a $R= 4.0 kpc$ circular orbit and place the dwarf on a circular orbit around the centre of the Milky Way at a radius of $R = 20.0 kpc$.  For all our simulations, we must also perform two comparison simulations in which the same cluster evovles in the dwarf potential only and the Milky Way potential only. The comparison clusters simulated in the Milky Way are always given a circular orbit at the same distance from the centre that the cluster will ultimately end up at.

\section{Results}

For clarity, in all figures we plot quantities corresponding to the cluster as it evolves in the combined potential, dwarf + Milky Way, in black. Quantities that represent cluster properties in the dwarf potential only and Milky Way potential only will always appear in red and blue respectively.

\subsection{Dwarf Falls into Milky Way}

\begin{figure} 
\caption{Simulation of a cluster around a $10^9 M_{\odot}$ dwarf that falls into the Milky Way with an initial half mass radius of $r_m = 3.2 pc$.}
\begin{subfigure}{0.5\textwidth}
\centering
\includegraphics[width=0.9\linewidth]{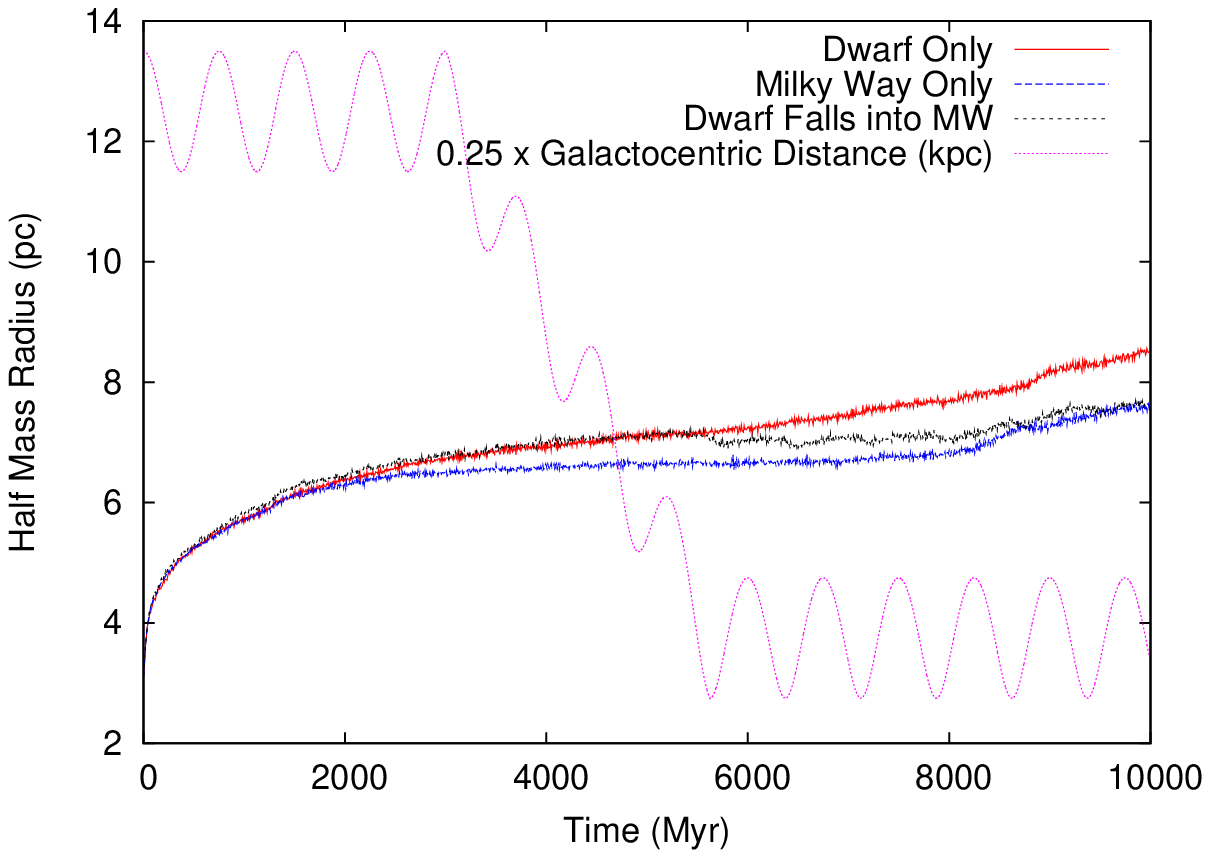}
\caption{Half mass radius in parsecs over time for the cluster in three potentials: dwarf only (red) at 4 kpc, Milky Way at 15 kpc (only) and in the combined potential of the dwarf falling into the Milky Way (black). The magenta line gives 0.25 x the distance between the cluster and the Milky Way centre in kiloparsecs.}
\label{fig:S15rad}
\end{subfigure}
\begin{subfigure}{.5\textwidth}
\centering
\includegraphics[width=0.9\linewidth]{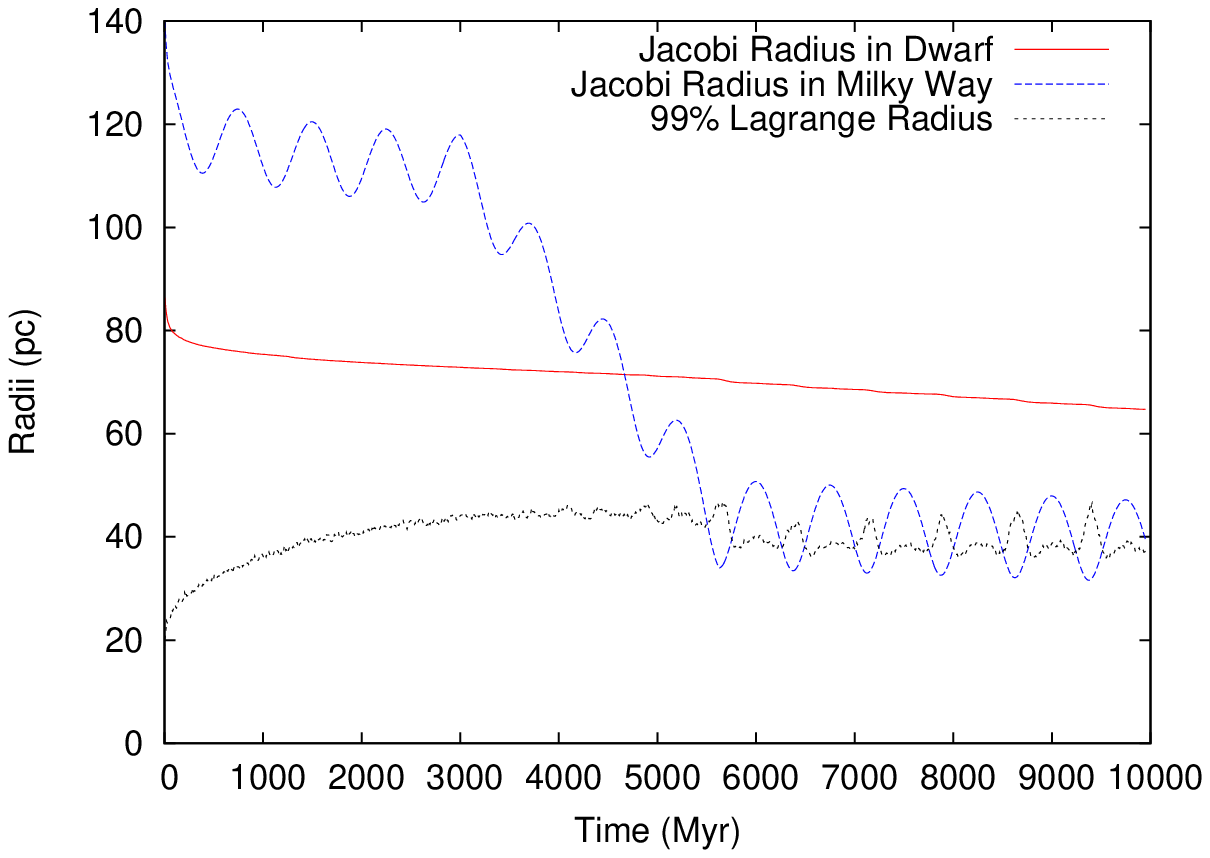}
\caption{The Jacobi radius of the cluster in the dwarf (red) and in the Milky Way (blue) as defined by Equation \ref{eq:Jacobi}. We also plot the cluster's 99 per cent Lagrange radius, $r_{99}$ in black.}
\label{fig:S15rj}
\end{subfigure}
\label{fig:S15}
\end{figure}

In our first simulation, we evolve the cluster on a circular orbit around a point mass dwarf ($M_D = 10^9 M_{\odot}$ and $R = 4.0 kpc$) which is located 50 kpc from the Milky Way centre. After 3 Gyr, we allow the distance between the dwarf and Milky Way to decrease until the separation between the two galaxies is 15 kpc. In Figure \ref{fig:S15rad}, we plot the half mass radius for a cluster in such a potential. For comparison, we also show clusters in the dwarf only and in the Milky Way only (orbiting on a circular orbit of $R = 15$ kpc). Additionally, to demonstrate how the cluster's position in the Milky Way potential changes over time, we plot the distance between the cluster and the Milky Way centre (multiply by 4 to get the correct value in kpc). The periodic variation in this distance is due to the cluster's circular orbit around the dwarf galaxy.

We can understand the cluster's size evolution in the combined potential by considering Figure \ref{fig:S15rj}, where we plot the Jacobi radius of the cluster in each of the individual galaxies as a function of time. We also plot in Figure \ref{fig:S15rj} the 99 \% Lagrange radius, $r_{99}$, a measure for the overall size of the cluster. The calculation of the Jacobi radius in each galaxy is done by considering the cluster's orbit through that potential and ignoring the effects of the other galaxy. Hence, to calculate $r_j$ for the cluster in the dwarf, we simply use the expression given in Equation \ref{eq:Jacobi} for circular orbits. However, no analytic expression for $r_j$ in the Milky Way exists for the particular orbit that we have used (effectively, a spiral inwards towards the Galactic centre). To obtain a sensible estimate for $r_j$ in this case, we use Equation 1 and evaluate it at the instantaneous position of the cluster in the Milky Way all along its orbit. The mass used in this equation is the mass enclosed by a circular orbit of radius equal to the position of the cluster in the Milky Way at any given time. Therefore, we see a decrease in $r_j$ for the Milky Way due to both the cluster's decreasing mass and decreasing Galactocentric distance in the Milky Way. The decrease in $r_j$ for the dwarf corresponds to the decreasing mass of the cluster only since the mass of the dwarf and radius of the cluster's orbit stays constant throughout the simulation. From this point in the text, the two values of Jacobi radius in the dwarf and Jacobi radius in the Milky Way will be abbreviated as $r_j^{D}$ and $r_j^{MW}$ respectively.

The evolution of the cluster's half mass radius in the combined potential is almost identical to its evolution in only the dwarf galaxy over the first several Gyr of the cluster's lifetime. This similarity indicates that when the dwarf is far from the Milky Way centre, the dwarf tides dominate the cluster. This idea is reinforced by Figure \ref{fig:S15rj} which shows that $r_j^{D}$ is much smaller than $r_j^{MW}$ during the first stage of the cluster's life. However, as the dwarf falls into the Milky Way, the $r_j^{MW}$ shrinks (the tides become stronger) and the Milky Way becomes the dominant galaxy in terms of tidal field strength. At about 5.5 Gyr (2.5 Gyr after the dwarf starts to fall into the Milky Way) the cluster's half mass radius suddenly starts to decrease. By examining Figure 1, we see that this sudden decrease corresponds to the time when $r_j^{MW}$ has decreased to a value such that $r_{99}$ becomes roughly equal to it. At this point, the cluster completely fills its Jacobi radius in the Milky Way and becomes very susceptible to tidal stripping. Also at this time, periodic bumps in the half mass radius emerge corresponding to the cluster passing closer to the Milky Way centre and then further away on its orbit around the dwarf galaxy. A short time after the dwarf reaches its final position in the Milky Way, at a separation of 15 kpc, the cluster's half mass radius completely overlaps with the half mass radius of the cluster that has evolved in the Milky Way only. This point of overlap corresponds to a physical time of about 8.0 Gyr. At the point in time when the cluster starts to respond to the Milky Way potential, the half mass relaxation time is approximately 1.3 Gyr. Hence, the cluster takes about 1.9 relaxation times to fully adjust to its new potential.

\subsubsection{Effect of Cluster Size}

\begin{figure} 
\caption{Simulation of a cluster around a $10^9 M_{\odot}$ dwarf that falls into the Milky Way with an initial half mass radius of $r_m = 4.0 pc$.}
\begin{subfigure}{0.5\textwidth}
\centering
\includegraphics[width=0.9\linewidth]{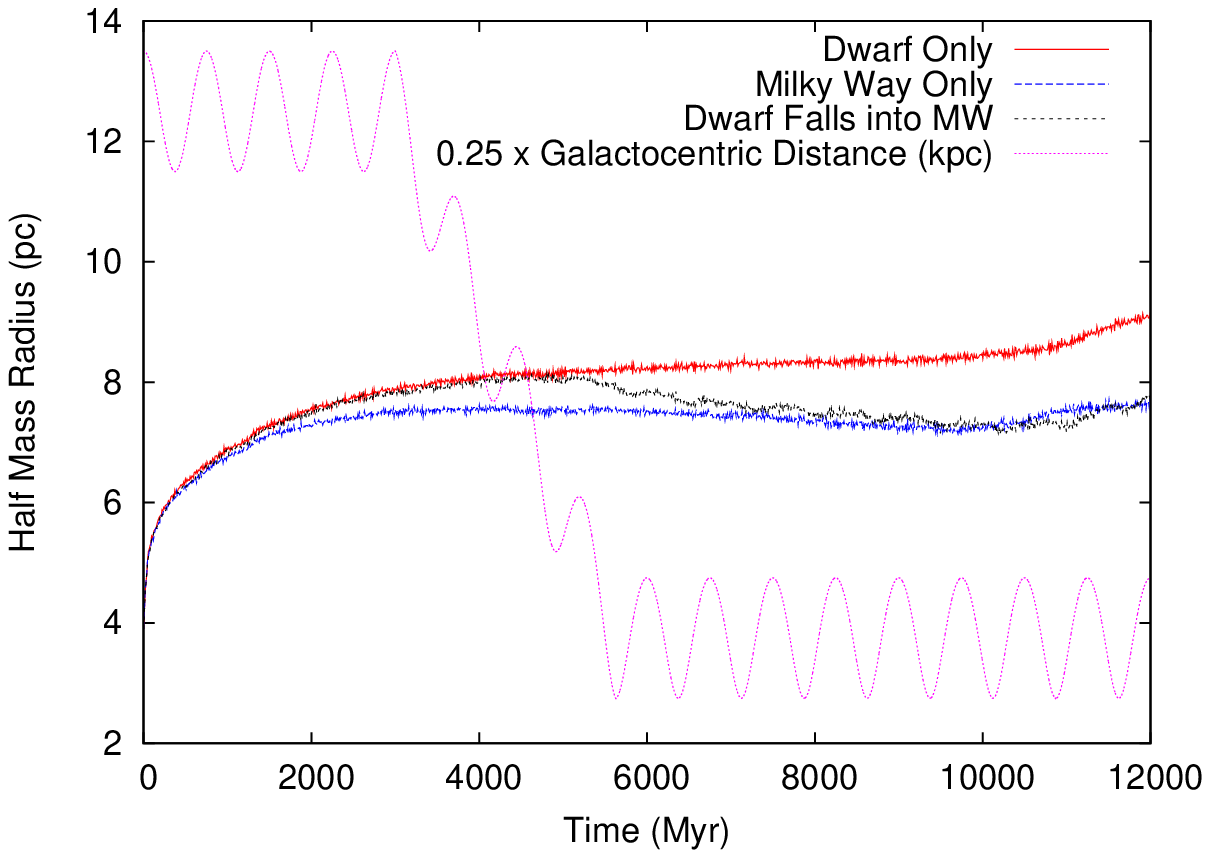}
\caption{Half mass radius over time in the three potentials as well as 0.25 x the Galactocentric distance in the Milky Way. Colours are the same as in Figure \ref{fig:S15rad}.}
\label{fig:S15HMR4rad}
\end{subfigure}
\begin{subfigure}{.5\textwidth}
\centering
\includegraphics[width=0.9\linewidth]{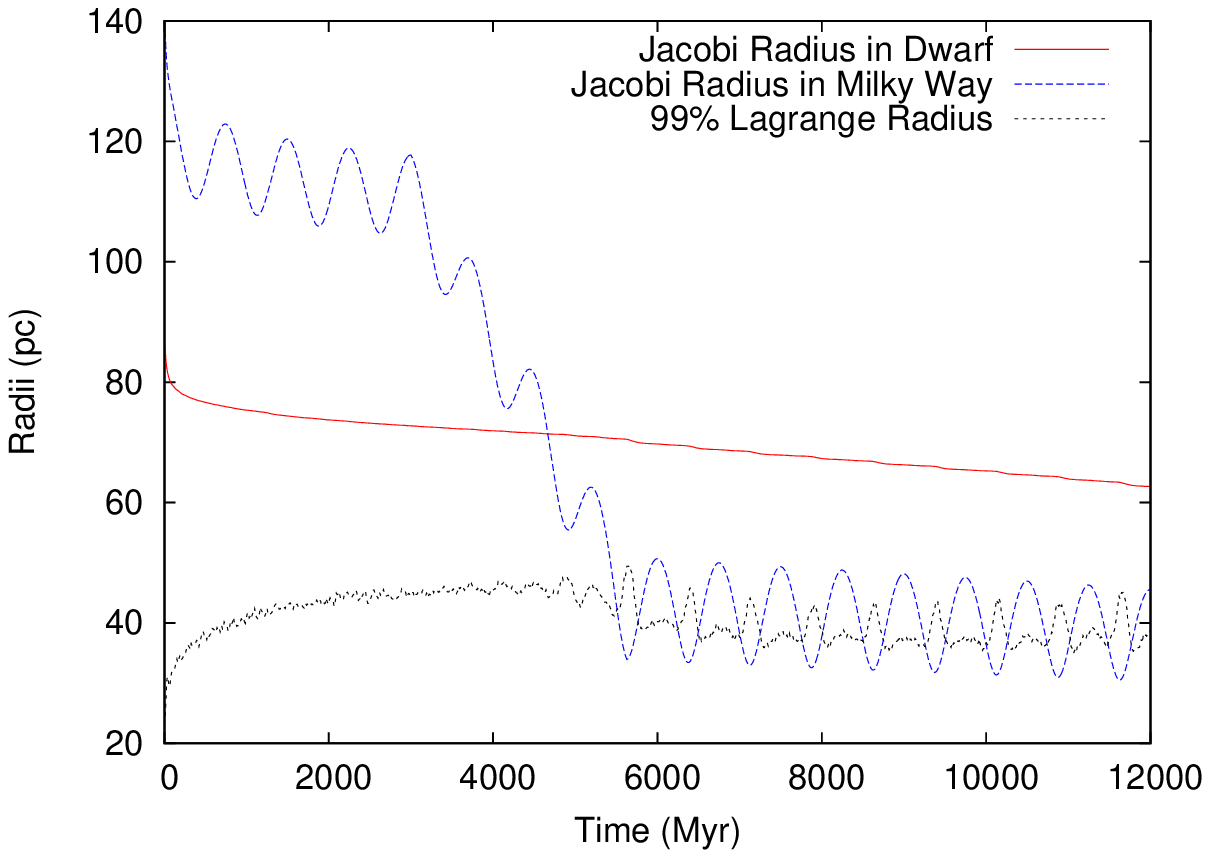}
\caption{The Jacobi radius in each potential and $r_{99}$ over time. Colours are the same as in Figure \ref{fig:S15rj}.}
\label{fig:S15HMR4rj}
\end{subfigure}
\label{fig:S15HMR4}
\end{figure}

To investigate how the initial size of the cluster affects our results, we perform another simulation with the same potential and larger initial half mass radius, $r_m = 4.0 pc$. The results of this simulation are plotted in Figure \ref{fig:S15HMR4}, showing the same information as Figure \ref{fig:S15}, discussed above. Comparing the evolution of the differently sized clusters, we see that they follow the same overall pattern: the cluster follows the same evolution it would have in the dwarf only early on, at some point the Milky Way tides begin to dominate and the cluster eventually has the same half mass radius as the cluster in the Milky Way only simulation. However, the half mass radius of the cluster with a larger initial size starts to decrease in response to the increasing tidal strength of the Milky Way before the the smaller cluster, at about 5.0 Gyr. Examining Figure \ref{fig:S15HMR4rj} reveals that this effect is due to the overlap of $r_{99}$ and $r_j^{MW}$ occurring at different times since differences in initial half mass radius correspond to differences in $r_{99}$.  

The larger cluster fully adjusts by 7.25 Gyr or about 1.4 relaxation times after the cluster starts to adjust (the half mass relaxation time at the cluster's first response is 1.6 Gyr).  Hence, in addition to the larger cluster starting to respond to the new potential earlier, it also takes a smaller amount of time (both in physical units and relaxation times) to complete its adjustment. Although bigger clusters have longer relaxation times they adjust more quickly because they are more vulnerable to the tidal field (their filling factors are larger). This result suggests that mass loss due to tidal stripping plays a more dominant role in changing the cluster's size than internal relaxation driven processes.

For clusters with smaller initial sizes, we expect the response to the Milky Way potential to begin at later times. However, it is conceivable that a cluster could be so small that, at the final position of the cluster in the Milky Way, $r_j^{MW}$ could still be larger than $r_{99}$. In this case, as long as $r_j^{MW}$ is smaller than $r_j^{D}$, the tides of the Milky Way will dominate and the cluster's size will be similar to the evolution of a Milky Way only cluster. However, in this case, the adjustment will not be as rapid since the tidal field of the Milky Way won't be able to strip stars as efficiently.

\subsubsection{Effect of a More Massive Dwarf}

\begin{figure}
\caption{Simulation of a cluster around a $10^{10} M_{\odot}$ dwarf that falls into the Milky Way.}
\begin{subfigure}{0.5\textwidth}
\centering
\includegraphics[width=0.9\linewidth]{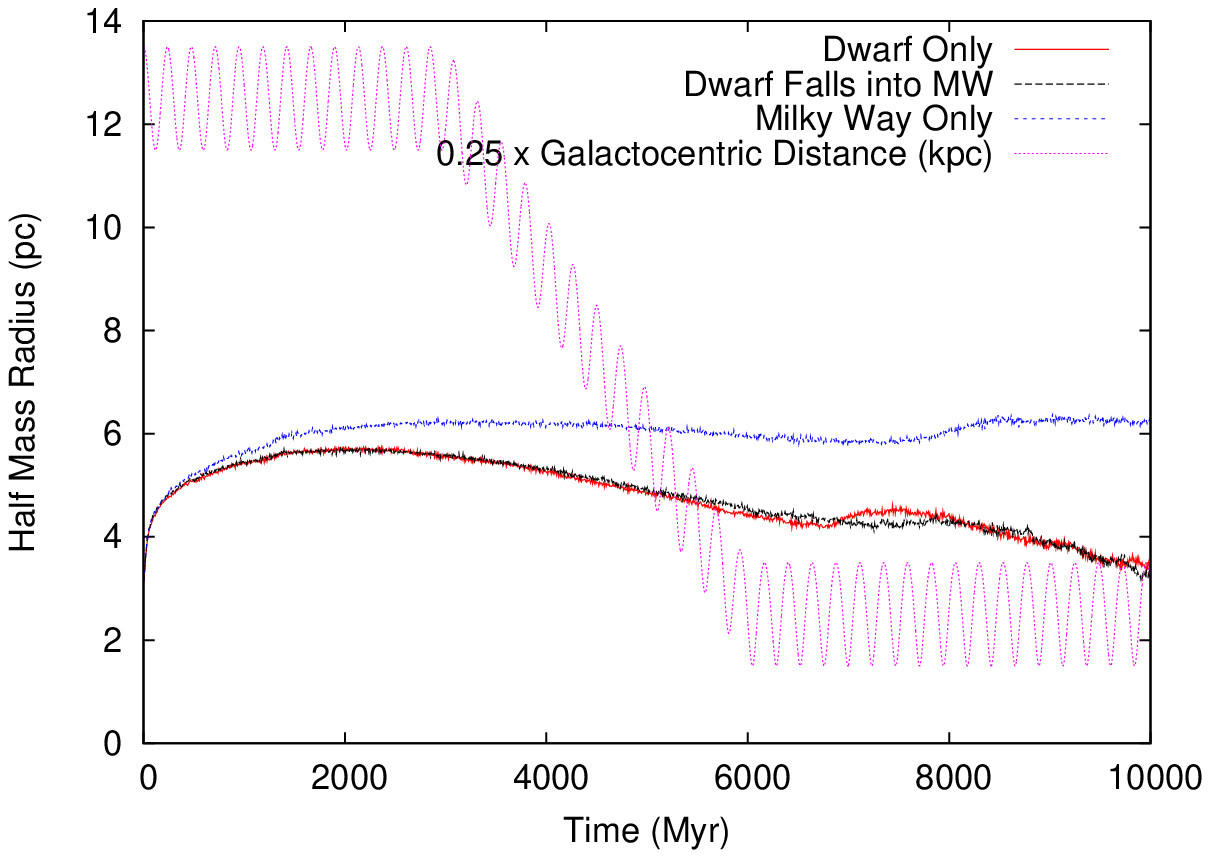}
\caption{Half mass radius over time in the three potentials as well as 0.25 x the Galactocentric distance in the Milky Way. Colours are the same as in Figure \ref{fig:S15rad}.}
\label{fig:S10DWLrad}
\end{subfigure}
\begin{subfigure}{.5\textwidth}
\centering
\includegraphics[width=0.9\linewidth]{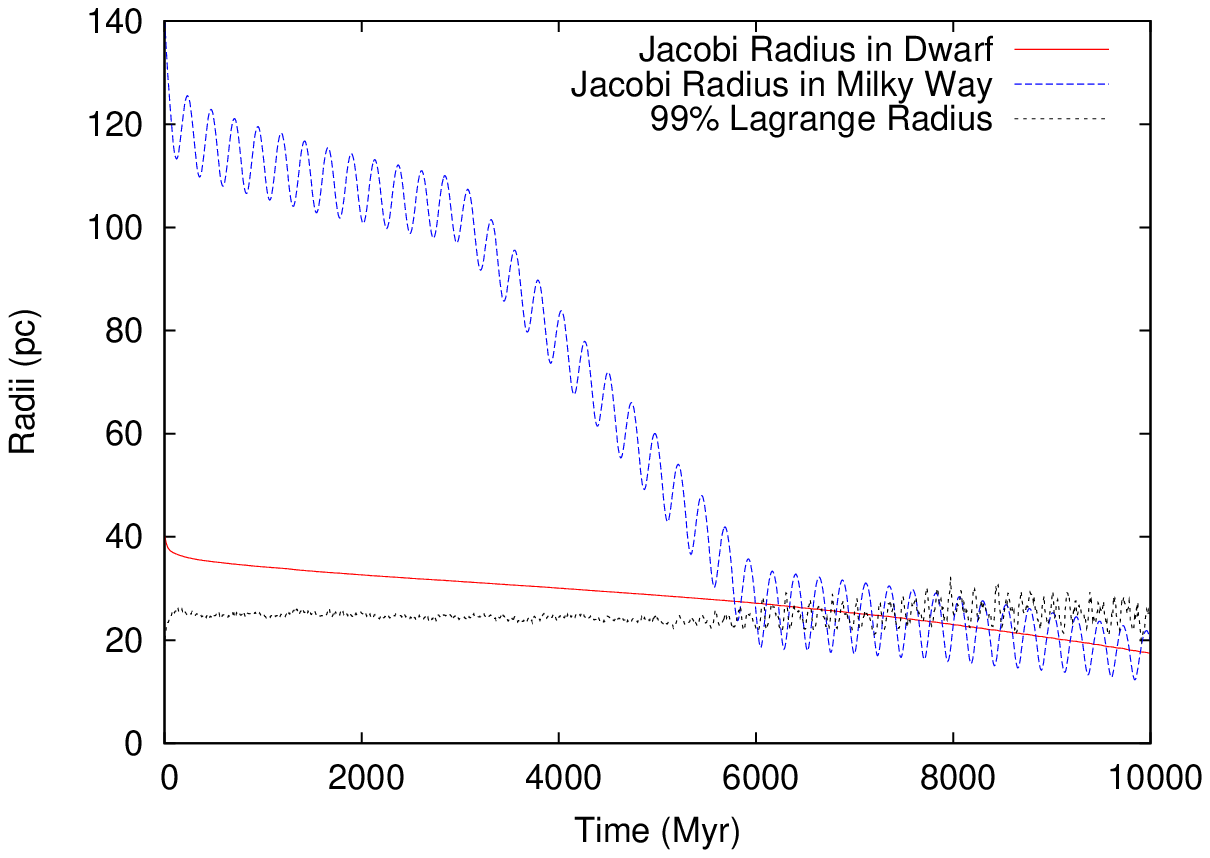}
\caption{The Jacobi radius in each potential and $r_{99}$ over time. Colours are the same as in Figure \ref{fig:S15rj}.}
\label{fig:S10DWLrj}
\end{subfigure}
\label{fig:S10DWL}
\end{figure}

We also study the effect of changing the dwarf's tidal field strength by performing a simulation with $M_D = 10^{10} M_{\odot}$. In this simulation, we allow the dwarf galaxy to fall into the Milky Way farther, to a separation of just 10 kpc. These parameters were chosen to make the final tidal field strengths of both galaxies roughly equal. This equality is demonstrated in Figure \ref{fig:S10DWLrj}, where we see that when the dwarf reaches its final position in the Milky Way, $r_j^{D}$ and the time averaged $r_j^{MW}$ are approximately equal.  In Figure \ref{fig:S10DWLrad} we again show the half mass radius of the system in the combined potential as well as in the dwarf and Milky Way alone. We see that the cluster seems largely unaffected by the Milky Way throughout its entire evolution, deviating only slightly from its evolution in the dwarf potential only. Essentially, the cluster behaves as though only one galaxy were present in setting its size. This effect occurs because the summation of the tidal forces from each galaxy is a vector summation. Since the cluster is orbiting around the dwarf galaxy, there will be phases of the cluster's orbit where the acceleration a star experiences towards the Milky Way is increased due to the presence of the dwarf galaxy (e.g. when the dwarf is between the cluster and the Milky Way). In this case, the instantaneous $r_j$ will be less than $r_J^MW = r_j^D$.   However, since this stage represents a small part of the cluster's orbit around the dwarf the time averaged $r_j$ will be close to the value of $r_j$ obtained by considering only one of the galaxies.

\subsection{Dwarf Evaporates}

\begin{figure} 
\caption{Simulation of a cluster orbiting around a $10^{10} M_{\odot}$ dwarf galaxy that evaporates as it orbits the Milky Way on a circular orbit of $R = 20.0 kpc$.}
\begin{subfigure}{0.5\textwidth}
\centering
\includegraphics[width=0.9\linewidth]{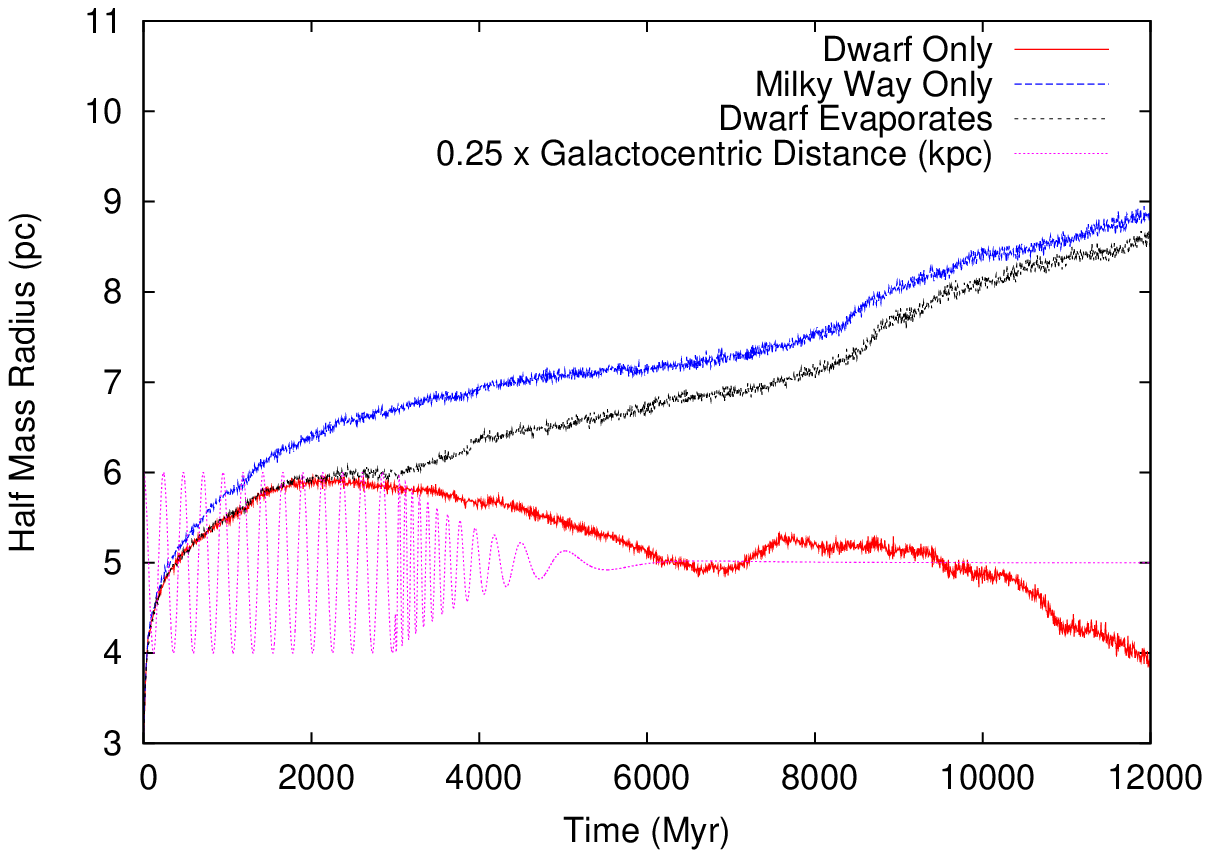}
\caption{Half mass radius over time in the three potentials as well as 0.25 x the Galactocentric distance in the Milky Way. Colours are the same as in Figure \ref{fig:S15rad}.}
\label{fig:DWLMW20rad}
\end{subfigure}
\begin{subfigure}{.5\textwidth}
\centering
\includegraphics[width=0.9\linewidth]{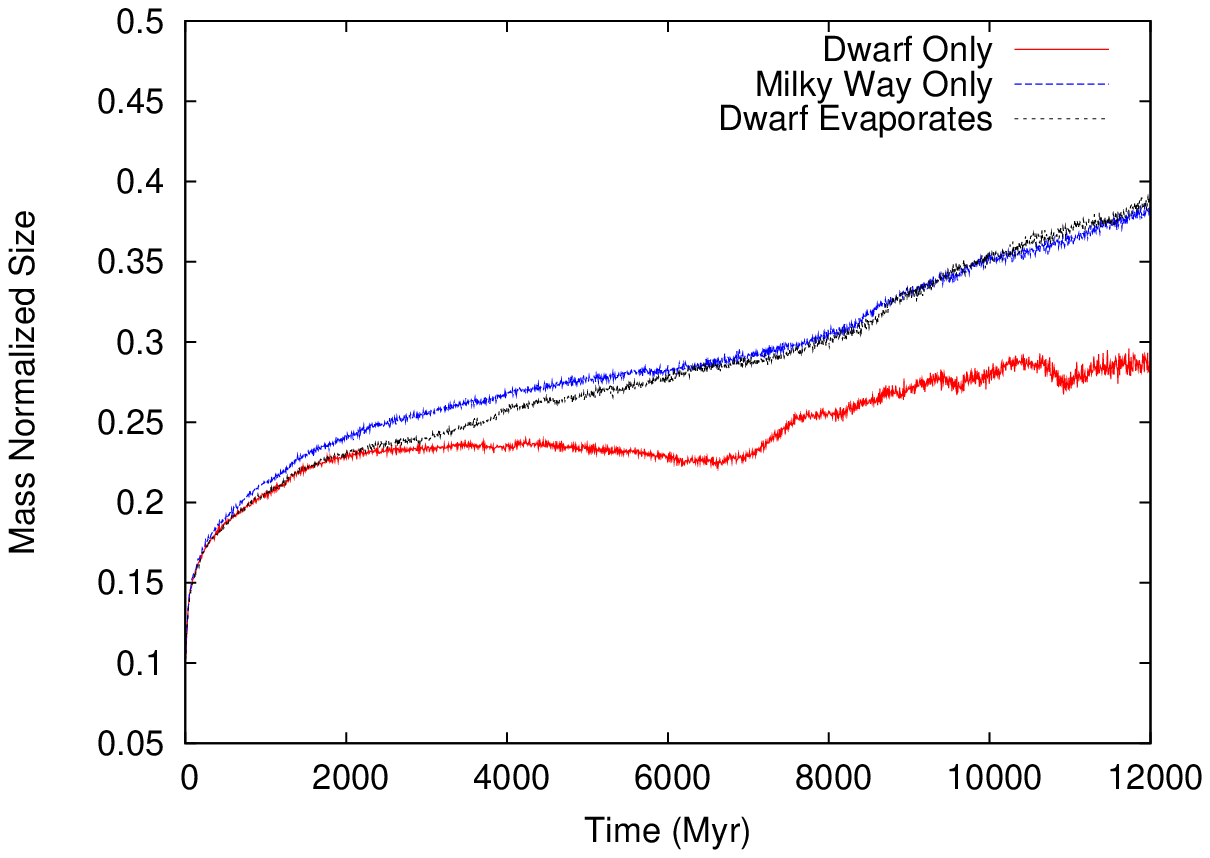}
\caption{Mass normalized size ($r_m/M^{(1/3)}$) of the cluster over time in the three potentials: dwarf only (red), Milky Way(blue), combined potential (black).}
\label{fig:DWLMW20mn}
\end{subfigure}
\begin{subfigure}{.5\textwidth}
\centering
\includegraphics[width=0.9\linewidth]{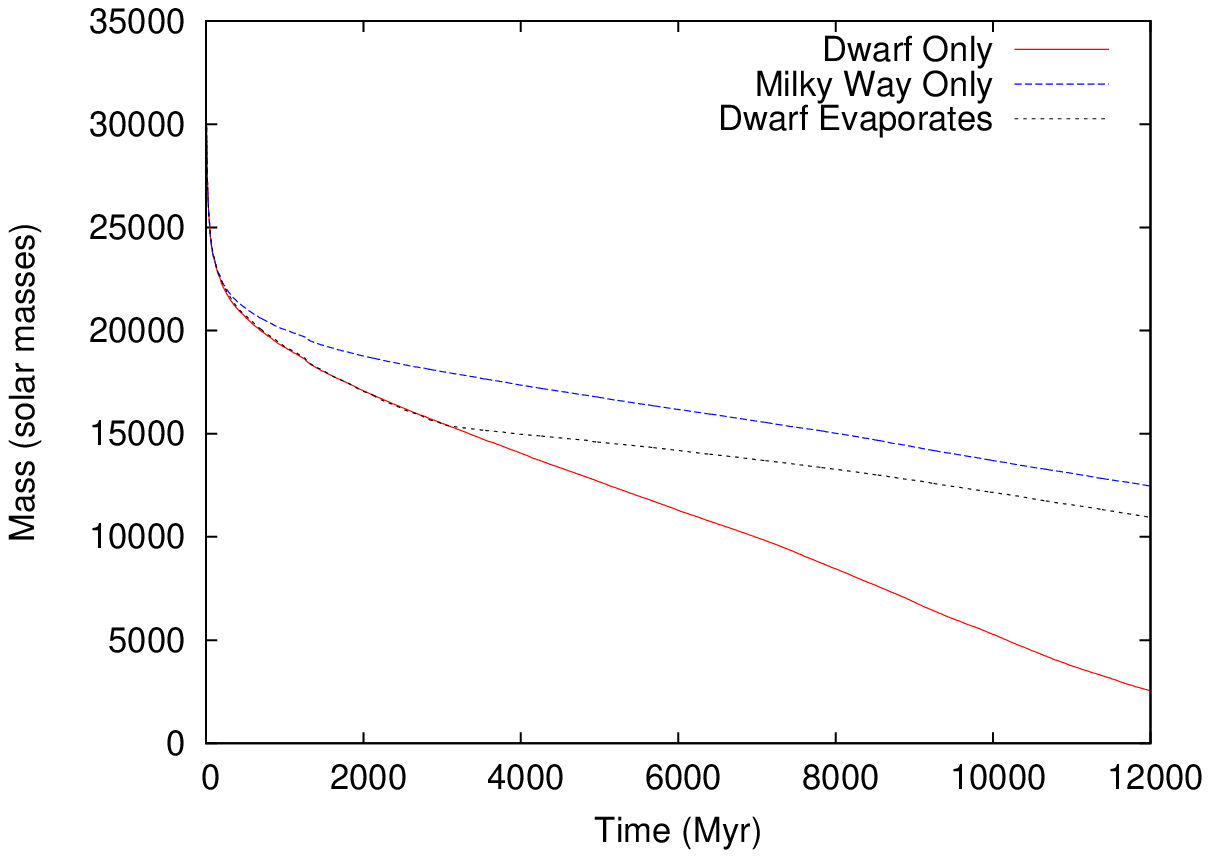}
\caption{Mass of the cluster in $M_{\odot}$ over time in the three potentials. Colours are the same as above.}
\label{fig:DWLMW20mass}
\end{subfigure}
\label{fig:DWLMW20}
\end{figure}

In our second scenario, we evolve a cluster on a circular orbit around a dwarf galaxy that in turn executes a circular orbit around the centre of the Milky Way. We use the same dwarf galaxy and orbit   ($M_D = 10^{10} M_{\odot}$, $R = 4.0 kpc$) as in the third ``dwarf falls" simulations presented above. The dwarf galaxy executes a circular orbit of radius $R = 20.0$ kpc around the centre of the Milky Way. After 3 Gyr of evolution in this system, we decrease the mass of the dwarf and radius of the cluster's orbit in the dwarf as described in Section 2, eventually leaving the cluster on a circular orbit in the Milky Way at 20.0 kpc.

In Figure \ref{fig:DWLMW20}, we show the cluster's half mass radius over time in this potential. Initially, the cluster follows an evolution similar to its evolution in the dwarf galaxy only, showing that the dwarf's tides dominate over the Milky Way tides in this configuration. When the mass of the dwarf begins to decrease, the tides weaken and we see an immediate increase in the cluster's half mass radius. By 6 Gyr, the dwarf's mass is effectively zero and the cluster is orbiting in the Milky Way only. The half mass radius of the cluster continues to expand and becomes very close to the half mass radius of the cluster living in the Milky Way its whole life.  To remove the effect of varying mass loss rates in different potentials, in Figure \ref{fig:DWLMW20mn}, we have shown the mass normalized radius ($r_m /M^{1/3}$) of the cluster in the three potentials. After the dwarf evaporates we see that mass normalized radius overlaps completely with the mass normalized radius for the cluster in the Milky Way potential only. Hence, the remaining difference in the clusters' half mass radii in these two potentials can be attributed to a difference in cluster mass, which we have plotted in Figure \ref{fig:DWLMW20mass}. 

We see in Figure \ref{fig:DWLMW20mass}, a large difference in the mass loss rate of the cluster in the Milky Way in comparison to the combined potential over the first 3 Gyr of evolution. Consequently, when the dwarf begins to evaporate and the cluster starts to adjust, a large difference in the masses of the clusters in the two potentials exists. For the size of the cluster to completely adjust the mass of the cluster must as well. Hence, the mass loss rate slows and the masses in the two potentials become comparable. However, the masses never completely overlap because the mass loss rate of the cluster in the combined potential cannot slow enough to compensate for such a large difference in cluster masses. This effect occurs when the cluster moves into a tidal field that is weaker than the original tidal field.  There is no such a effect when the cluster moves from a weak tidal field to a strong one, as in the ``dwarf falls" case shown in Figure \ref{fig:S15}. In this case, an acceleration of the mass loss rate is driving the evolution of the cluster, rather than a deceleration. The increased mass loss rate acts quickly to bring the cluster's mass down to what it would have been had it been living in the stronger tidal field its whole life. 

In the simulation presented in Figure \ref{fig:DWLMW20}, we start with a dwarf galaxy that has a stronger tidal field at $R = 4.0 kpc$ than the Milky Way. To examine the effects of starting the cluster in a dwarf galaxy that has a weaker tidal field than the Milky Way, we performed additional simulations in which we decreased the mass of the dwarf and let it orbit at a distance closer to the centre of the Milky Way. Before the dwarf evaporates, it has little effect on the cluster because the Milky Way's tidal field is stronger. The orbit of the cluster in the Milky Way (passing closer and farther away from the centre due to the orbit around the dwarf) does not have an effect on the cluster's size. Even though the tidal field of the Milky Way varies quite a bit on such an orbit, the cluster appears to evolve according to the mean tidal field strength of the Milky Way along that path. The dwarf evaporating decreases its tidal field strength even more and so the cluster continues to have a half mass radius as dictated by the Milky Way's tidal field. 

\section{Summary and Discussion}

In our simulations, we simulate a cluster that orbits around a dwarf galaxy which either falls into the Milky Way or evaporates and leaves the cluster orbiting in the Milky Way only. The size of a cluster is determined by whichever galaxy has the strongest tidal field strength. Hence, in the ``dwarf falls" simulations, when the separation between the dwarf-cluster pair and Milky Way is such that the dwarf tidal field strength is stronger (has a smaller $r_j$), the cluster has the same size as it would have if it evolved solely in the dwarf. Conversely, when the dwarf is close enough to the Milky Way centre such that the Milky Way's tides begin to dominate, the cluster size decreases and eventually becomes the same size as a cluster which spends its entire lifetime in the Milky Way. This adjustment occurs because the stronger tidal field of the Milky Way can strip stars closer to centre of the cluster, keeping the cluster confined to a region where the tides from the dwarf are not strong enough to strip stars form the cluster. The transition between these two regimes occurs when $r_j^{MW}$ becomes comparable to $r_{99}$, i.e. the physical limit of the cluster. In the ``dwarf evaporates" simulations, the cluster responds to the weakening tidal field of the dwarf galaxy immediately, deviating from the size it would have in the non-evaporating dwarf. As the dwarf tides go to zero, the Milky Way begins to dominate and the subsequent size evolution is determined by the Milky Way tides entirely, leaving the cluster with a similar size as a cluster that has evolved completely in the Milky Way only. Any remaining size difference can be attributed to slightly different masses since the mass-normalized size of the cluster in the combined potential and the Milky Way converge almost immediately after the dwarf has evaporated. 

Taking all the results from the above simulations together, we can construct the full picture of a star cluster's evolution inside a dwarf galaxy-Milky Way merger. The two scenarios used for the gravitational potential on the cluster, ``dwarf falls" and ``dwarf evaporates" represent the key processes affecting a cluster in this type of galaxy merger. In particular, in the merger these processes will happen simultaneously since as the separation between the two galaxies diminishes, the Milky Way will strip mass away from the dwarf. These processes have the same net effect, to increase the Milky Way's tidal field strength with respect to that of the dwarf. The simulations we perform in this work show that a cluster will have a size determined by whichever tidal field is the strongest at any one point. Whenever the Milky Way's tides take over, whether its due to the dwarf evaporating, the cluster becoming close to the Milky Way centre or both, the cluster will respond quickly to the new potential and ultimately become the same size as a cluster that has evolved solely in the Milky Way. 

The results of this work are consistent with our previous findings detailed in \citet{Miholics14}. In that paper, we used an instantaneous change in galactic potential, from dwarf galaxy to Milky Way. Using this method, we found the same result as in this work: clusters adjust quickly in response to the new Milky Way potential, such that their size appear as if they had always been orbiting in the Milky Way on that particular orbit. The main difference between this and our previous work are the timescales over which adjustments occur. Naturally, since the potential experienced by the cluster changes gradually over time in the simulations presented here, the changes in the cluster's half mass radius take longer to occur. However, once the potential reaches its final configuration and the tides on the cluster no longer vary, the cluster generally adjusts quickly. 

For the changes in potential modelled here, the size adjustment takes place within $\approx$ 1-2 current cluster half mass relaxation times. We found that clusters with a larger initial size take less time to adjust to their new potential than their smaller counterparts. Such differences in adjustment time exist because larger clusters are more susceptible to stripping by the tidal field. This effect occurs despite the longer relaxation times of larger clusters, suggesting that the new tidal filling factor of the cluster plays the largest role in determining the adjustment timescale. Hence, even  the most extended globular clusters in the Milky Way with long relaxation times would have adjusted quickly (if they were accreted) as long as they were not too tidally underfilling. We also note that altering the relative field strengths (initial versus final), may slightly alter the timescale over which adjustments occur.

Our work is also consistent with recent simulations performed by \citet{Bianchini15}. These simulations are similar to our ``dwarf evaporates" case but in their case after the dwarf evaporates the cluster is left in isolation, rather than in the Milky Way. They also found that the cluster would respond to the change in potential by expanding rapidly, achieving sizes similar to (but always slightly less than) the size of a completely isolated cluster. The results of these simulations in conjuction with our work show that clusters accreted from dwarf galaxies should not be larger than clusters born in the Milky Way on the same orbit. In particular, the extended nature of some observed clusters in the Milky Way \citep{Harris96} is not a consequence of being accreted from a dwarf galaxy and experiencing a change in tides. Instead, the large size of these clusters most likely reflects some difference in their initial conditions at birth such as a large initial size \citep{Zonoozi11,Zonoozi14}. \citet{Bianchini15} also investigate the effect of turning the dwarf off instantaneously versus smoothly over time and reach similar conclusions as we do when comparing our current work with our previous work, as discussed above.

Our results suggest that it should not be possible to determine the origin of a cluster in the Milky Way based on its size alone. In particular, the distributions of tidally fillling and underfilling clusters in the Milky Way should not be affected by the presence of clusters accreted by dwarf galaxies (assuming that initial cluster sizes born in dwarf galaxies are similar to those born in the Milky Way). This result is consistent with the findings of \citet{Baumgardt10}. Although, they found evidence for two distinct populations of clusters in the Milky Way, one filling and one underfilling, they found no correlation between these groups and the group of clusters that are expected to be accreted from dwarf galaxies based on other properties.

An interesting extension of the work presented here would be to study a cluster inside a dwarf that is on an elliptical orbit around the Milky Way. In this case, the galaxy with the dominant tidal field strength might change along the orbit. If these oscillations occurred on a long timescale, our results suggest that the cluster's size would also oscillate between its size in the dwarf galaxy only and the Milky Way only, quickly adjusting to whatever tidal field was the strongest. However, if these oscillations happened on a short timescale then the cluster would probably exhibit a size as determined by the average tidal field strength of the two galaxies. It is worth noting that such a situation would probably be short lived since the dwarf would have to be quite close to the centre of the Milky Way for this to occur and the cluster would be stripped away from the dwarf quickly.

A possible next step in this work is to perform a galaxy simulation of a merger between a dwarf and the Milky Way and extract from it the potential to use in the $N$-body simulation. This method would allow one to study the orbits of clusters inside the combined potential of the two galaxies (such as the elliptical ones discussed in the previous paragraph) as well as help to quantify the relative timing of the two processes studied here (falling and evaporating). Since our models probe the key processes important to cluster evolution in galaxy mergers, this method should yield similar results on the sizes of most accreted clusters. However, using a galaxy simulation would allow for the study of accreted clusters with more dramatic tidal fields than the ones studied here (such as strong encounters with the Milky Way's disc, dark matter substructure or other clusters) and may lead to the discovery of important physical processes that affect some accreted clusters and not others.

\section*{Acknowledgements}

We would like to thank Florent Renaud greatly for his always quick and helpful assistance in our use of \textsc{Nbody6tt}. We would also like to thank the referee for providing constructive comments and helping to increase the level of this work.

We acknowledge the use of SHARCnet resources in the completion of this work. M.M., A.S. and J.W. are supported by NSERC (Natural Sciences and Engineering Research Council of Canada).








\bsp	
\label{lastpage}
\end{document}